\documentclass[twocolumn,nofootinbib]{revtex4-1}   

\usepackage{amsfonts}  

\newcommand{\ket}[1]{|#1\rangle}
\newcommand{\braket}[2]{\langle #1|#2\rangle}

\begin{document}

\title{Quantum mechanics: last stop for reductionism}
\author{Gabriele Carcassi}
 \affiliation{Brookhaven National Laboratory, Upton, NY 11973}
 \email{carcassi@bnl.gov}
\date{February 29, 2012}

\begin{abstract}
The state space of a homogeneous body is derived under two different assumptions: infinitesimal reducibility and irreducibility. The first assumption leads to a real vector space, used in classical mechanics, while the second one leads to a complex vector space, used in quantum mechanics.
\end{abstract}

\maketitle

Reductionism, the study of complex things by reducing them to the interaction of their parts, is such a fundamental idea in science that it's seldom explicitly discussed. I'd like to focus on the subject for a moment, and present this simple, clear-cut distinction: classical mechanics is suitable for objects we assume to be infinitesimally reducible, we can picture them as made of infinitesimally tiny pieces interacting with each other; while quantum mechanics is suitable for objects that we assume to be irreducible, we cannot picture them as made up of smaller pieces, every interaction is with the whole.

I will focus only on the state space, as the features most specific to quantum mechanics (e.g. interference, uncertainty principle, entanglement) can be found just from the state definition, and many other concepts (e.g. equations of motion, lagrangian and hamiltonian mechanics) are essentially the same just re-cast on different definitions of state spaces\cite{dirac}\cite{neumann}.

\section{Homogeneous system}

We will consider a homogeneous system, where each part exhibits the same physical properties. Let's define $\mathbb{S}$ as the set of all possible states $\mathbf{s}$. While we still have no idea what this set looks like, we can start to define the following operations:

\textbf{Magnitude}: $\mathcal{M}(\mathbf{s} \in \mathbb{S}) \Rightarrow \mathbb{R}$. Given a state, returns a scalar value that indicates the size of the state. Quantities like mass or charge will be proportional to this number.

\textbf{Resize}: $\mathcal{R}(a \in \mathbb{R}, \mathbf{s} \in \mathbb{S}) \Rightarrow \mathbf{s'} \in \mathbb{S} : \mathcal{M}(\mathbf{s'}) = a \mathcal{M}(\mathbf{s})$. Given a state and a scalar, gives a new state where the magnitude and all the extensive properties are increased by a factor of $a$, while all the intensive properties remain the same. That is, it scales the size by a factor $a$.

\textbf{Group}: $\mathcal{G}(\mathbf{s}_1, \mathbf{s}_2 \in \mathbb{S}) \Rightarrow \mathbf{s} \in \mathbb{S}\otimes\mathbb{S}$. This operations allows us to put states together to form complex systems. It returns a state that can be reduced to the original ones. The two states are parts of the new state, which is of a different type: it's the state of two systems ($\mathbb{S}\otimes\mathbb{S}$), not one ($\mathbb{S}$).

\textbf{Combine}: $\mathcal{C}(\mathbf{s}_1, \mathbf{s}_2 \in \mathbb{S}) \Rightarrow \mathbf{s} \in \mathbb{S}$. This operation allows us to modify states, which is essential if we want to have operations like time evolution. Given two states, returns a third which is a modified version of the first based on the "instructions" given by the second. The final state is in the same space as the arguments.

We now ask: how does the magnitude of a combined state relate to the magnitude of the original states? It turns out that this alone will be extremely useful to characterize the state space.

\section{Infinitesimal reducibility or classical mechanics}

Let's assume that our system is infinitesimally reducible.\footnote{Reducibility should not be confused with divisibility. For example, a unicellular organism is \emph{divisible into} two unicellular organism (through mitosis) but is \emph{not reducible to} them (a cell is not made of two cells). A magnet, instead, is reducible to a north and a south pole, but is not divisible into a north and a south pole (you'd get a north and a south pole for each part). In quantum mechanics, a high energy photon is divisible into a electron-positron pair (it can decay into them) but is not reducible to it (it's not made of other particles). A proton is reducible to quark and gluons, but is not divisible into them (you'd get showers of other particles).} That is, any state $\mathbf{s} \in \mathbb{S}$ can be reduced to two states $\mathbf{s}_1$ and $\mathbf{s}_2$, which in turn and can be reduced, and so on.

Since we can reduce and re-group the state however we want, grouping and combining are essentially the same operation: $\mathcal{C}(\mathbf{s}_1, \mathbf{s}_2) = \mathbf{s} \Rightarrow \mathbf{s}_1 \otimes \mathbf{s}_2 = \mathcal{G} (\mathbf{s}_1, \mathbf{s}_2)$. In other words, we can always study the states that we are combining as independent parts. The magnitude of a composed state will then be the combined magnitude of the individual states: $\mathcal{M}(\mathcal{C}(\mathbf{s}_1, \mathbf{s}_2)) = \mathcal{M}(\mathbf{s}_1) + \mathcal{M}(\mathbf{s}_2)$; the mass/charge/etc...~of a system is the sum of the mass/charge/etc...~of all its parts. \emph{The magnitude is linear under combination and resize.}

We can reduce each state into parts that have unique evolution under all circumstances. Given that position and momentum identify unique evolution, we can divide the space using these as parameters, and each state will be reassembled from these. Putting all of this together, we can set:

\textbf{State space}: each state is described by a vector in a real inner product space.

\textbf{Group/Combine}: $\mathcal{G}(\mathbf{s}_1, \mathbf{s}_2) = \mathcal{C}(\mathbf{s}_1, \mathbf{s}_2) =  \mathbf{s}_1 + \mathbf{s}_2$. Grouping and combining are the sum of the vectors.

\textbf{Resize}: $\mathcal{R}(a, \mathbf{s}) = a \mathbf{s} $. Resizing is the multiplication by the resizing constant.

\textbf{Basis}: $\mathbf{s} = \int\int_{\mathbf{x}, \mathbf{p}} \rho(\mathbf{x}, \mathbf{p}) \mathbf{e}_{\mathbf{x}, \mathbf{p}} \mathbf{dx dp}$. The natural base $\mathbf{e}_{\mathbf{x}, \mathbf{p}}$ is made from the states that identify unique evolution, parameterized by position and momentum. All states are derived from the basis through combining and resizing. $\rho(\mathbf{x}, \mathbf{p}) = \mathbf{s} \cdot \mathbf{e}_{\mathbf{x}, \mathbf{p}}$ is the magnitude density for each point of state (or phase) space.

\textbf{Magnitude}: $\mathcal{M}(\mathbf{s}) = \int\int_{\mathbf{x}, \mathbf{p}} \mathbf{s} \cdot \mathbf{e}_{\mathbf{x}, \mathbf{p}} \mathbf{dx dp}$. For the total magnitude, we sum up all the contributions from each basis.\footnote{Note that the magnitude does not correspond to the norm of the vector, as the norm adds each contribution quadratically instead of linearly.}

To recap: we assume the system is infinitesimally reducible, we reduce each state into those parts that have unique evolution by using the position and momentum parametrization (or any other you may need) and we keep track of the magnitude of each part.

\section{Irreducibility or quantum mechanics}

Now for the opposite assumption: the system is irreducible. This means we can't track the motion of any of its parts independently and determine the inner dynamics. Our description must be, so to speak, determined as a whole and uncertain on the details.\footnote{This uncertainty leaks out in terms of the uncertainty principle: if the whole state were specified by a single value for both position and momentum, it would mean that all its parts would also be perfectly specified, we would be able to track the motion of each part: the system would not be irreducible at all!} We are not going to try to fully characterize this uncertainty except for two things: it must exist and, if the system is homogeneous, each part must carry an equal share. That is: the total uncertainty must be proportional to the magnitude. So, instead of asking directly how the magnitude changes under combination, we can look at how the uncertainty changes.

Given its stochastic nature, the combined uncertainty will not be a simple sum: some parts may be correlated (will interfere constructively), some may be anti-correlated (will interfere destructively) and some may be uncorrelated (won't interfere). If we expect this to change as the statistical variance, we will have $\mathcal{M}(\mathcal{C}(\mathbf{s}_1, \mathbf{s}_2)) = \mathcal{M}(\mathbf{s}_1) + \mathcal{M}(\mathbf{s}_2) + 2 \sqrt{\mathcal{M}(\mathbf{s}_1)\mathcal{M}(\mathbf{s}_2)} \mathit{Corr}(\mathbf{s}_1, \mathbf{s}_2)$ where $-1 \leq \mathit{Corr}(\mathbf{s}_1, \mathbf{s}_2) \leq +1$ is the correlation of the uncertainty between the two states. The expression can be rewritten as $\mathcal{M}(\mathcal{C}(\mathbf{s}_1, \mathbf{s}_2)) = (\sqrt{\mathcal{M}(\mathbf{s}_1)} + e^{-\imath \theta} \sqrt{\mathcal{M}(\mathbf{s}_2)} ) (\sqrt{\mathcal{M}(\mathbf{s}_1)} + e^{+\imath \theta} \sqrt{\mathcal{M}(\mathbf{s}_2)} )$ where $\theta = \arccos{(\mathit{Corr})}$ we can call the correlation angle. This shows that, while the magnitude does not combine linearly, the square root of the magnitude does. With this in mind, we can set:

\textbf{State space}: each state is described by a vector in a complex inner product space.

\textbf{Group}: $\mathcal{G}(\mathbf{s}_1, \mathbf{s}_2) = \ket{s_1} \ket{s_2}$. Grouping is the product of the two vectors.

\textbf{Combine}: $\mathcal{A}(\mathbf{s}_1, \mathbf{s}_2) = \ket{s_1} + \ket{s_2}$. Combining is the sum.

\textbf{Resize}: $\mathcal{R}(a, \mathbf{s}) = \sqrt{a} \ket{s} $. Resizing is the multiplication by the square root of the resizing constant. Multiplication by a phase, instead, $e^{\imath b} \ket{s}$ will change the correlation between this and all other states.

\textbf{Basis}: $\ket{s} = \int_{\mathbf{x}} \rho_x(\mathbf{x}) e^{\imath \theta_x(\mathbf{x})} \ket{x} \mathbf{dx}$. We can only choose one parametrization at a time (e.g. position or momentum).  $\rho_x(\mathbf{x}) e^{\imath \theta_x(\mathbf{x})} = \braket{x}{s}$ where $\rho_x$ is the square root of the magnitude that is at position $x$ while $\theta_x$ is the correlation angle relative to the base.

\textbf{Magnitude}: $\mathcal{M}(\mathbf{s}) = \braket{s}{s} = \int_{\mathbf{x}} \braket{s}{x} \braket{x}{s} \mathbf{dx}$. The magnitude corresponds to the norm of the vector. \footnote{If a homogeneous system is reducible to a number of systems that cannot be further reduced, one would expect these final systems to be of the same magnitude. So, by convention, one may require actual physically possible states to be unitary $\braket{s}{s} = 1$.}

To recap: we assume the system is irreducible, any inner dynamics is therefore unspecified, this uncertainty and its correlation is what produces interference when combining states. In each base, we keep track of the square root of the magnitude and what the correlation is.

\section{Final remarks}

Deriving the rest of the theory is possible, but would naturally require a much longer work. Here I just wanted to show how quickly we can go from abstract concepts to the underpinning of both classical and quantum mechanics. And how this highlights the similarities (homogeneity, operations defined) and the differences (reducibility, linearity of $\mathcal{M}$ or of $\sqrt{\mathcal{M}}$) in hopefully simpler and more intuitive terms than other works.

\end{document}